# Hydrogen transport within graphene multilayers by means of flexural phonons

Vito Dario Camiola[a,b], Riccardo Farchioni[b,c], Vittorio Pellegrini[b,d], Valentina Tozzini[a,b]

[a]*Istituto Nanoscienze, Cnr, Piazza San Silvestro 12, 56127 Pisa, Italy*
[b]*NEST- Scuola Normale Superiore, Piazza San Silvestro 12, 56127 Pisa, Italy*
[c]*Dipartimento di Fisica 'E. Fermi', Università di Pisa Largo B. Pontecorvo 3-56127 Pisa, Italy,*
[d]*Graphene Labs, iit Istituto Italiano di Tecnologia, Via Morego, 30 16163 Genova, Italy*

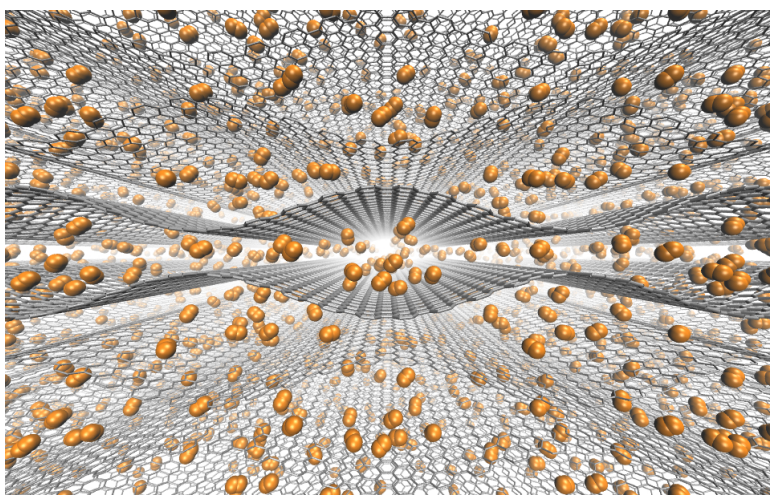

**Abstract**

Graphene sustains transverse out-of-plane mechanical vibrations (flexural phonons). At the nanometer scale, these appear as travelling ripples, or cavities, if excited in counter-phase in alternate multilayers. In this work we explore by means of classical molecular dynamics simulations the possibility of using these moving nano-cavities to actively transport hydrogen. We find that the gas can be efficiently transported for hundreds of nanometers in the wave propagation direction, before the phonons damp down. Therefore, this effect could be used to move and pump gases through multilayers graphene based frameworks.

# 1. Introduction

Exploring new perspectives for the green technologies is one of the challenges of the third millennium, in which the need for un-polluting and renewable powering has become primary. In this context, the use of hydrogen as a fuel is promising, since the energy released in its combustion is three times that of the average hydrocarbons and the released product is water[1]. However, gaseous or liquid hydrogen in tanks poses problems of safety, cost and unpractical conditions for transportation and storage[2]. This produced a great effort in the search for solid-state systems and materials for H storage[3,4]. Graphene has most of the characteristics that could make it an ideal system[5]: it is very robust and resistant, light-weight and with a large surface to mass ratio, which potentially increases its gravimetric capacity.

Interactions of hydrogen with graphene are generally classified as "chemical" or "physical". The first easily occurs with atomic hydrogen, producing a stable hydrogen adsorbate on graphene, which locally changes its hybridization from sp2 to sp3 forming C-H bonds (chemisorption)[6,7] However, molecular hydrogen has a high barrier for chemisorption, which requires the use of some catalytic strategy[8,9]. Consequently, using this process for hydrogen storage implies slow adsorption and desorption kinetics at room temperature, or working at high temperatures, which is a problem in common with of other solids state storage systems, like metal hydrides[10]. On the other hand, $H_2$ is easily physisorbed onto graphene, but the binding energy is very low, relying on van der Waals (vdW) interactions[11]. It was theoretically shown[11] and confirmed by a large number of experiments[12] that gravimetric densities (GD) as large as 6-8% can be reached within nano-porous or multi-layered graphene at cryogenic temperatures and/or high pressures, suggesting niche applications such as large storage facilities, or powering in extreme conditions where e.g. conventional batteries are useless. However, the upper limit at room temperature is estimated around 2%[11,12], the same order of, but not better than, conventional energy storage systems. Thus graphene based H storage systems could find applications in standard conditions provided they perform better in terms of safety, robustness and practical use.

The extended 2D nature of graphene and its unique mechanical properties offer alternative strategies to improve storage performances. For instance, graphene can sustain transverse out of plane vibrations, namely ZA or flexural phonons[13], which behave as "traveling ripples". According to calculated and measured dispersion relations[14], if the wavelength (i.e. ripples separation) is in the nm scale, the corresponding wave frequency is in the THz range. In our previous work[15], we demonstrated by Density Functional Theory based simulations that coherent ZA phonons in this wavelength range are effective for detaching chemisorbed hydrogen on graphene at room temperature, and also to reduce the chemisorption barrier[7], indicating a route to obtain fast loading/release kinetics in standard conditions.

In this work, we investigate whether flexural phonons could also be exploited in the case of physisorption. In particular, we study by molecular dynamics simulations the possibility of using coherent ZA phonons to transport physisorbed $H_2$ through multilayered nanostructures. We include hydrogen into the nanometer sized cavities formed within multilayers in the presence of ZA phonons of nanometric wavelength and ~THz frequency, and simulate the dynamics of the system on the hundreds of ns time scale by means of classical molecular dynamics with empirical Force Fields. Our main aim is to show that active hydrogen transportation is produced, due to the hydrogen tendency to accumulate within concavities and driven by the coherent vibrations of the multilayer. Overall the effect is similar to the nano-pumping demonstrated in carbon nanotubes[16]. However, here it is realized in a 3D system (a graphene multilayer) rather than in a 1D one, offering the possibility of massive transportation on macroscopic scales, yet controllable at the nanometric level, besides the access to more strategies for the activation and sustain of the necessary coherent vibrations.

After the description of the model systems, we report results of simulations for single layer, bilayer and multilayers, with and without hydrogen, and finally illustrate conclusions and possible perspectives for using these results in practical applications.

## 2. System and Methods

The simulations were performed within the classical molecular dynamics framework, using empirical Force Fields (FF) for the interactions between atoms. The carbon-carbon interactions were described by the Tersoff FF[17], whose functional form (reported in Appendix) includes three body interactions and automatically accounts for the hybridization via the dependence on the bond order. Several versions of the Tersoff FF are available with slightly different parameterization optimized for different purposes. We chose here the one suitable for the reproduction of the phonon dispersion relations of graphene[14].

The model system (represented in Fig 1) is an approximately squared graphene sheet of 4-4.5nm side (See Appendix for details) with periodic boundary conditions in the in-plane directions (x, corresponding to the harmchair edge, and y, the zig-zag edge). This system was simulated as an isolated monolayer, as a bi-layer, or as a multilayer, repeating the bi-layer periodically in the z direction. The inter-layer distance and the z periodicity are chosen in such a way that the minimal distance between layers in the presence of out-of-plane deformations is always larger than ~4.5 Å, so to minimize the direct inter-layer interactions.

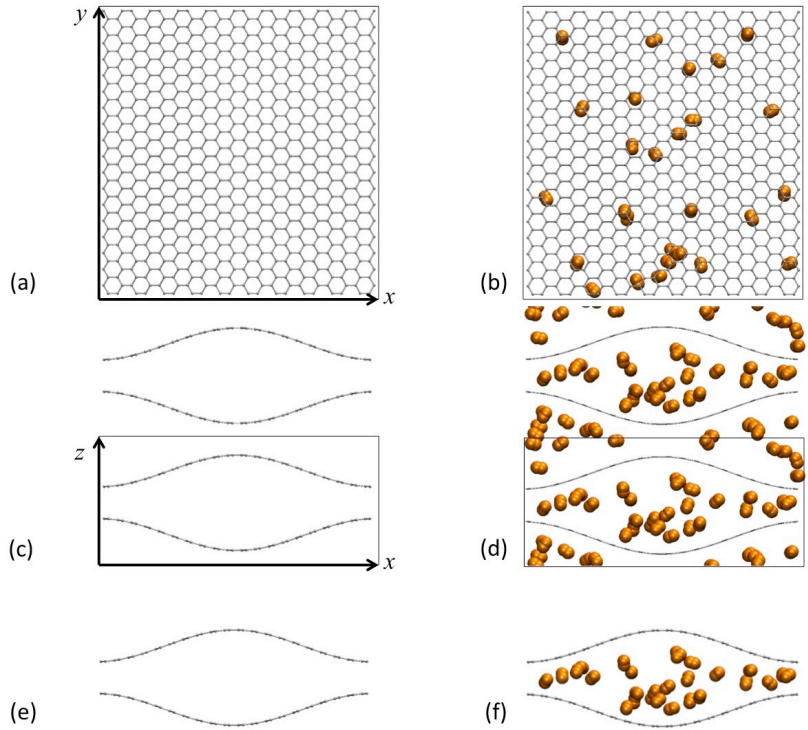

**Figure 1.** Illustration of the model systems. (a) Top view of the supercell graphene only. x and y are indicated. (b) Same as (a), with hydrogen included. (c) Side view of the multilayer graphene supercell, with z and y axes indicated. One periodic image along z is reported. (d) Same as (c) with hydrogen included in all the cavities. (e) Side view of the bylayer. (f) Same as (e) with hydrogen included into the cavity.

ZA phonons are generated in this system in x and y directions as sinusoidal out of plane displacements with the same periodicity of the super-cell, which, therefore, determines the wavelength of the phonon. The latter and the phonon amplitude $A$ were chosen as a compromise between the need of enclosing cavities of nanometer size within couples of layers ad the need of having near-room temperatures (the phonon temperature scales as $T \propto (\nu A)^2$, and the dependence of $\nu$ on $\lambda$ for these phonons is as $1/\lambda^2$). The final choice is a super-cell (and wavelength) of ~4nm and the phonon amplitude $A=2.5$Å (see the Appendix for details). The resulting frequency was re-determined from simulations, and seen to remain in the THz range, though affected by anharmonicity and by interactions within the system (see Table 1). The resonant frequency of stationary waves was used to generate a starting velocity field with a ¼ wave dephasing with respect to the displacement, to generate the traveling wave, corresponding to a coherent phonon.

Graphene sheets were simulated alone and in the presence of molecular hydrogen, included in the cavity formed by the bilayer and in the multiple cavities of the multilayer (see Fig 1, and Table 1 for the complete list of simulations). The $H_2$ molecule is represented as a rigid rod, while the C – H and H – H intermolecular interactions are represented by Lennard-Jones potentials (See Table A.1). The volume of the system is maintained constant. $H_2$ is added at room temperature and at the pressure of ~100atm. Such a high pressure was chosen here to amplify the effects, while a detailed study of the behavior of the system in the range down to ambient pressure will be published in a forthcoming study[18]. The starting configuration and velocity of the added hydrogen were taken from a previous simulation of a thermalized pure hydrogen box.

The Newton equations of motions are integrated with the Velocity Verlet algorithm[19] with a timestep of 1 fs. Vibrational spectra were evaluated by the Fourier transform of the velocity autocorrelation function[20]. Simulations were performed within the microcanonical ensemble, and implemented within the DL_POLY_classic software package. Software tools to create the input and to analyze the output (e.g. the vibrational spectra) were programmed in house.

## 3. Results and discussion

We first performed simulations of the monolayer, bilayer and multi-layer systems without hydrogen, with ZA phonons (sim 1-8 in Table 1). In the bilayer and multilayer cases the nearby layers are initialized to vibrate with a half-period dephasing (counterphase). This creates a "traveling cavity" between subsequent layers. The vibrational spectra are reported in Fig 2, and show sharp peaks at the phonon frequency, with small variation between each other and with respect to the stationary waves of the monolayer of the same wavelength and amplitude (see Table 1). The frequency shows also little dependence on the direction (x or y), in agreement with the fact that at these wavelengths the dispersion relations are nearly isotropic[17]. The independence of the spectral main component on the layers number (mono, bi or multi-layers) indicates that with the chosen interlayer distances, the inter layers interactions are negligible, and do not produce appreciable phonon broadening. As a matter of fact no phonon damping is observed during the simulation time. However, approaching the ns timescale, in the multilayer system the sheets tend to dephase, and some spurious vibrations appear, signatures of the emergence of the expected and natural loss of coherency. We remark that at these large amplitudes we expect to be out of the harmonic regime[21].

Table 1. Simulation list, with input parametes. Output quantities are reported in italics.

| Sim # | System | k dir | wave type | A (Å) | $\nu$ (THz) | $H_2$ Press (atm) | T (K) | Run length (ps) |
|---|---|---|---|---|---|---|---|---|
| 1 | monolayer | x | stationary | 2.50 | *0.77* | – | *222* | 300 |
| 2 | monolayer | y | stationary | 2.50 | *0.70* | – | *208* | 300 |
| 3 | monolayer | x | traveling | 2.50 | *0.84* | – | *410* | 300 |
| 4 | monolayer | y | traveling | 2.50 | *0.84* | – | *476* | 300 |
| 5 | bilayer | x | traveling | 2.50 | *0.84* | – | *425* | 300 |
| 6 | bilayer | y | traveling | 2.50 | *0.78* | – | *422* | 300 |
| 7 | multilayer | x | traveling | 2.50 | *0.84* | – | *421* | 300 |
| 8 | multilayer | y | traveling | 2.50 | *0.79* | – | *393* | 300 |
| 9 | bilayer | x | traveling | 2.50 | *0.79* | 100 | *444* | 300 |
| 10 | bilayer | y | traveling | 2.50 | *0.78* | 100 | *392* | 300 |
| 11 | multilayer | x | traveling | 2.50 | *0.67* | 100 | *381* | 300 |
| 12 | multilayer | y | traveling | 2.50 | *0.72* | 100 | *343* | 300 |

The net result is that the cavities formed by subsequent sheets travels together with the phonons at the velocity of $v=\nu\lambda \sim 3.6$ nm/ps = $3.6 \times 10^3$ m/s. The dependence of these values on the supercell size and on the phonon amplitude will be investigated in a forthcoming paper, together with the effect of anharmonicity and of the specific FF chosen[22]. In this work, we focus in the reaction of the system to the inclusion of molecular hydrogen.

$H_2$ was then included into the cavity of the bilayer or within all the cavities in the multilayer case, and the system was let evolve. Fig 3 reports the results for the multilayer with phonons traveling in the x (a,c,e) and y (b,d,f) directions. The first observation is the thermalization of the system as an effect of the interaction between hydrogen and graphene (panels a, b, top). An analysis of the average velocities of $H_2$ molecules in the three directions (panels a,b, middle, colored lines), sheds light on how the energy interchange between the graphene and the gas occurs: the gas immediately gains a net velocity (3.5-4 nm/ps) in the propagation direction of the phonon, while the motion remains random-walk like in the other in plane direction, and oscillating due to bumps with the sheets in the z directions (black lines). This means that the phonon transfers net momentum to hydrogen, namely, active $H_2$ transport is realized. As a matter of fact, during the 300ps of simulation, the average path traveled by hydrogen is ~400 nm in the case of phonon propagating in x direction (harmchair), little less in the case of phonon propagating in the y direction (zig-zag, see panels a,b, bottom; movies illustrating the simulations are also reported as Supporting Information). The distance traveled in the other directions is negligible, as expected.

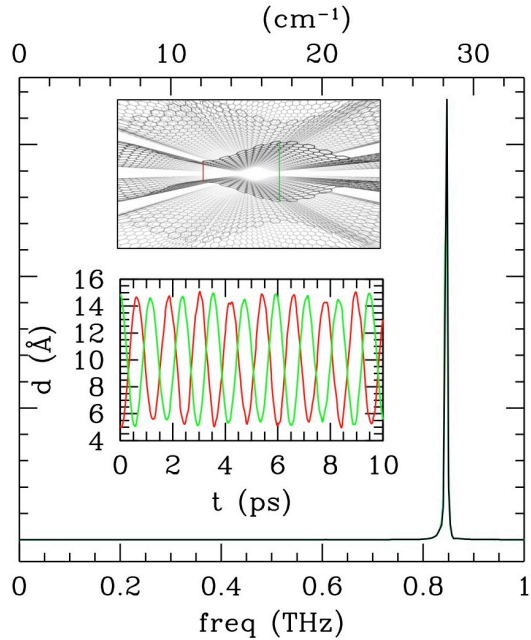

**Figure 2.** Vibrational spectrum evaluated on the 300ps simulation of the dynamics of the multilayer with the ZA phonons superimposed. The inset plot reports the evolution in time of the interlayer distances indicate in the reported structure with corresponding colors.

Fig 3 also shows that the process is very efficient at the beginning, during the first 100ps. In the second half of the simulation, the $H_2$ average velocity decreases rapidly. An inspection of the trajectory shows that the transport efficiency decreases as the relative phase and coherency of the phonons is lost (Fig 3 panels c,d): at the last 100 ns of the simulations, the sheets move in phase, with a half period de-phasing with respect to the starting configuration, implying that the nano-sized moving cavities are destroyed. In addition, the phonon amplitude and the frequency are smaller with respect to the starting ones. These changes are also testified by the broadening of the vibrational spectra of the multilayer towards the lower frequencies (Fig 3, panels e,f). Finally, the relative distance of the sheets changes, and they tend to form bilayers with single layers at the distance of ~7Å, each bilayer separated by ~13 Å. Incidentally, in static and flat multilayers the interlayer distance of ~7Å was theoretically shown to be the optimal one at which the vdW forces cooperate and produce the maximum $H_2$ intake[11]. It seems, therefore, that the system spontaneously adjusts in the optimal configuration to maximize the interaction with hydrogen, if it is let free to evolve. However this configuration is not the optimal one for the active gas transportation. This is not surprising: the system spontaneously evolves towards the thermodynamic equilibrium, while active transport is an out of equilibrium process. Clearly, the out of equilibrium condition requires energy to be maintained, namely some kind of powering. Simulations including external forces emulating powering are currently in progress and will be included in a forthcoming work.

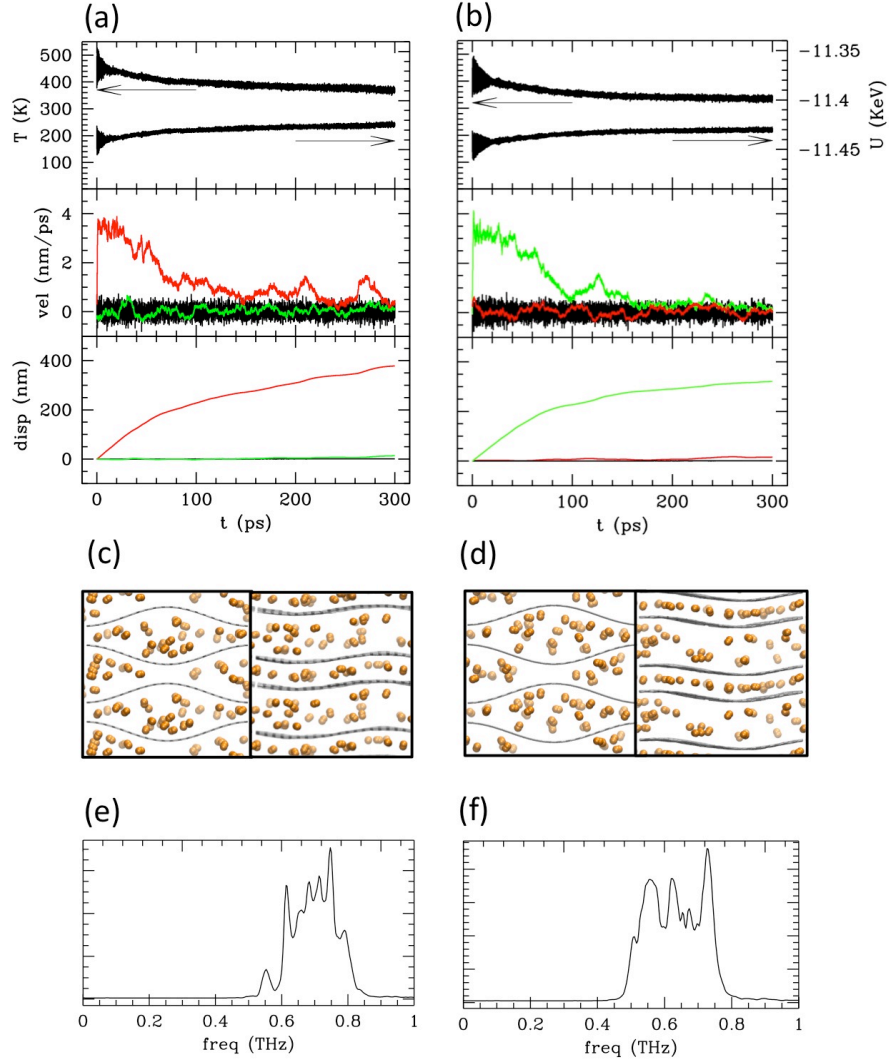

**Figure 3.** Simulations of $H_2$ at P~100 atm physisorbed in multilayers with ZA phonons traveling in x directions (harmchair direction, panels on the left, a,c,e, simulation 11) and in y directions (zig-zag direction, panels on the right, b,d,f, simulation 12). Panels a,b, report, from top to bottom: the temperature and (scale on the left) potential energy of the system (scale on the right) of the system during simulation; the average velocity components of $H_2$ (x red, y green, z black); the distance traveled by $H_2$ in the three directions (same color code). Panels (c) and (c): starting (left) and final (right) configuration of the system, lateral view from y direction (c) and from x direction (d). Panels (e) and (f): vibrational spectra of the graphene sheets during simulation.

The analysis of simulations 9 and 10 of the bilayers with hydrogen (not shown) reveals a similar behavior at the beginning: the average velocity of hydrogen in the propagation direction suddenly increases at ~3.5 ns/ps. However, during the first ~60ps the graphene sheets separate quite fast, due to high hydrogen pressure, which, in this case, is not counterbalanced by the presence of other layers. In spite of the sheet separation, the momentum transferred from graphene to hydrogen in the starting phase persists during the simulation, and a net transport of hydrogen of ~1μm is observed during the 300ps of simulation, although the sheet separation produces dispersion of hydrogen over a larger y-z section and consequently a reduction of the flux.

We can summarize these results in the following main points: (i) ZA coherent phonons can easily transfer momentum to the gas in the direction of the wave propagation, and create a macroscopic flux and net $H_2$ transportation (ii) as an effect of the gas pressure, the sheets tend to separate. In the case of multilayers large spatial separation is hindered, thus the system tends to form periodic bilayers with interlayer distance of ~7Å, which optimizes the hydrogen mediated vdW inter-layers interactions. (iii) These interactions soon produce phonon damping and dephasing, with consequent loss of efficiency of the gas transportation. The system, however, enters a regime in which coherent phonons are still present, though with lower amplitude, different phase and frequency. Therefore a non-null macroscopic flux is still present in the propagation direction, within the simulation time. It is likely that the phonons finally dump down in the ns timescale, but the average distance traveled by $H_2$ molecules after in the first 300ps is already ~400 nm, thus it is likely that even in this case they can almost reach the μm scale before complete phonon damping.

## 4. Conclusions and perspectives

This work illustrates by classical molecular dynamics simulations the possibility of molecular hydrogen transportation through graphene multilayers by means of coherent flexural phonons with wavelength in the nanometer scale and frequency in the THz range, and ~Å sized amplitude. ZA Phonons are excited only at the beginning of the simulation. The system is filled with hydrogen, and then let freely evolve.

The simulations show that vibrationally excited multilayers are able to transport hydrogen on the μm scale in less than 1 ns. In fact, the average hydrogen velocity in the optimal conditions is of the order of the phonon phase velocity. The absence of any phonon maintaining mechanism and the very high hydrogen pressure produce sheets separation and phonon damping and de-phasing within the ns scale, with consequent loss of efficiency of the mechanism. This is expected, because no powering is included in the simulation. At the ambient pressure, it is likely that the damping process will be proportionally slower, and the hydrogen could be transported for much longer distances. On the other hand at smaller pressures the flux will be proportionally smaller, leading to a comparable net amount of transported hydrogen.

This work provides the proof of concept that flexural phonons could be exploited to force flux of hydrogen through graphene sheets, i.e. act as a gas nanopump. A similar mechanism was previously proposed in carbon nanotubes, activated by Rayleigh traveling waves[23]. However, at variance with nanotubes, graphene multilayers give the possibility of volumetric gas transportation on the nano to micro scale. In addition, the graphene sheets offer more flexibility in the modulation of the phonon frequency and amplitude, allowing more room for the optimization of the process.

This effect could be used, for instance, to increase the internal pressure of physisorbed hydrogen, to fill vessels or other hydrogen storage means, e.g. graphene-based 3D frameworks, and, finally, to improve the intake/release phases in a graphene-based hydrogen storage device. At the same time, the mechanism here proposed could provide the nano-scale control of the gas (or even fluid) motion and direction through graphene networks. This might have applications in the field of advanced catalysis[24].

In any case, a strategy to excite and maintain coherent flexural phonons and the optimal interlayer distance is necessary. These aspects can be rather straightforwardly included in simulations, by means of constraints and external fields[25], and are the matter of a forthcoming paper. In principle, the multilayer specific structure and geometry offers different strategies to realize ZA phonons excitation, such as lateral strain[26], anchoring the sheets to a piezoelectric substrate, or functionalizing the multi-layers with optically active spacers. The fact that the phonons frequencies are in the THz range also suggests the possibility of exploiting coherent radiation sources in that range[27]. Investigations on the electromechanical coupling of graphene (bare or functionalized) are necessary to this aim, which are currently in the course[28]. Clearly, experimental verification of the effects demonstrated in these simulations would be beneficial, and we hope that this work could be a stimulus to that aim.

**Ackwnowledgements**

We acknowledge financial support by the EU, 7$^{th}$ FP, Graphene Flagship (contract no. NECT-ICT-604391) and the "Compunet platform" of Istituto Italiano di Tecnologia.

**Appendix: Force Field and simulations details**

The Tersoff FF has the following functional form:

$$V(r_{ij}) = f_c(r_{ij})\left[af_R(r_{ij}) - b_{ij}f_A(r_{ij})\right]$$

with

$$f_R(r) = A\exp(-l_1 r)$$

$$f_A(r) = B\exp(-l_2 r)$$

$$a = 1$$

$$b_{ij} = (1 + b^n \zeta_{ij}^n)^{-1/2n}$$

$$\zeta_{ij} = \sum_k f_c(r_{ij}) g(\theta_{ijk}) \exp([l_3(r_{ij} - r_{ijk})]^3)$$

$$g(\theta_{ijk}) = 1 + \frac{c^2}{d^2} - \frac{c^2}{d^2 + (h - \cos\theta_{ijk})^2}$$

$$f_c(r) = \begin{cases} 1 & r < S - D \\ \frac{1}{2}\left(1 - \sin\left(\frac{\pi(r-S)}{2D}\right)\right) & S - D < r < S + D \\ 0 & r > S + D \end{cases}$$

We used the parameterization from ref [17] reported in Table A.1.

**Table A.1.** Model system, and Force Field parameters.

| Force Field Parameters | | | | |
|---|---|---|---|---|
| **Interaction** | **FF** | **Parameters** | | |
| **C-C** | Tersoff | $l_1$=3.488Å$^{-1}$ $l_2$=2.211Å$^{-1}$ $l_3$=0.0Å$^{-1}$ R=1.95Å D=0.15Å S=1.95Å | A=139.6 eV B=430.0 eV | n=0.7275 c=38050 b=1.572×10$^{-7}$ d=4.348 h=-0.930 |
| **C-H** | LJ | σ= 2.99Å | ε= 0.002 eV | |
| **H-H** | LJ | σ = 2.72Å | ε=0.001 eV | |
| **H-H bond** | restrain | l=0.74 Å | | |
| **Super-cell parameters** | | | | |
| | **x (Å)** | **y (Å)** | **z (Å)** | **Atom #** |
| monolayer | 43.17 | 44.86 | 60.00 | 720 C |
| bilayer | 43.17 | 44.86 | 250.0 | 1440 C, 54(x) 58(y) H |
| multilayer | 43.17 | 44.86 | 20.00 | 1440 C, 98(x) 96(y) H |